\documentclass[prd,showpacs,aps]{revtex4}
\usepackage{amssymb}
\usepackage{amsmath}
\usepackage{epsfig}
\usepackage{latexsym}
\usepackage{hyperref}
\usepackage{graphicx}
\usepackage{subfigure}
\newcommand{\ml}{Margolus-Levitin Theorem }
\newcommand{\mll}{Margolus-Levitin Theorem}
\begin{document}
\title{Covariant versions of \mll}
\author{Qiaojun Cao}\email{23cqj@163.com}
\author{Yi-Xin Chen}\email{yxchen@zimp.zju.edu.cn}
\author{Jian-Long Li}\email{marryrene@gmail.com}
\affiliation{Zhejiang Institute of Modern Physics, Zhejiang
University, Hangzhou 310027, China}

\begin{abstract}
The \ml is a limitation on the minimal evolving time of a quantum
system from its initial state to its orthogonal state. It also
supplies a bound on the maximal operations or events can occur
within a volume of spacetime. We compare \ml with other form of
limitation on minimal evolving time,  and present covariant versions
of \ml in special and general relativity respectively. The relation
among entropy bound, maximum information flow and computational
limit are discussed, by applying the covariant versions of \mll.
\end{abstract}
 \maketitle

\section{Introduction}
In the view of quantum theory, the time-energy uncertainty relation
is interpreted as the limitation of how fast can a quantum system
evolve from its initial state to an distinguishable state,
equivalently, to an orthogonal state. In this relation, $\Delta t$
can be understood as the minimal evolving time between two distinct
states of the quantum system. The time-energy uncertainty relation
$\Delta t \geqslant \pi \hbar/2 \Delta E$ indicates the minimal
evolving time is constrained by the energy spread of the system.
Recently, Margolus and Levitin gave an other interpretation of
Heisenberg principle. In their proposition, the average energy
instead of the energy spread of the system is constraining the
minimal evolving time $\Delta t$. The \ml has several advantages.
For an isolated system the average energy is conserved while the
energy spread can be arbitrary. More over, for complicated system
the average energy is easy to determined while the energy spread
require more information or assumption from the system. As a result,
the \ml is widely used on a series of
topics\cite{ML,lloyd1,lloyd2,lloyd3,lloyd4,lloyd5,Zych,Ng,Xiao}.
\\

Some direct applications of \ml are taken to evaluate the speed of
ultimate quantum computer\cite{lloyd1}, the total operations
performed by the universe\cite{lloyd2} and the accuracy of measuring
the spacetime geometry\cite{lloyd3,Ng,Xiao}. Following \ml , once
the average energy of the system is determined, the minimal evolving
time between two distinct states is also determined. A quantum
computer with every operation performed as quickly as the minimal
evolving time has reached the maximum speed of the average energy.
If we perceive the universe as such an quantum computer, as we
already know the energy density and the age scale of the universe,
we can work out the total operations performed by the universe from
its beginning. This lead us to the third application, what is the
accuracy of measuring the spacetime geometry? A spacetime region can
be viewed as a container filled with operations or events. The
accuracy of spacetime measurement depends on density and
distribution of the operations or events. Accordingly, these
applications of \ml are using abstract or concrete models, including
quantum mechanical, cosmological, and quantum gravitational ones.
Since the \ml is quantum mechanical, and is deduced in pure state
system with a non-interacting hamiltonian. What is the validity of
\ml on these applications?
\\

In some previous work, Lloyd and his cooperators have discussed the
extension of \ml in composite systems, mixed states, and entangled
systems\cite{lloyd4,lloyd5}. They showed the \ml is still valid to
estimate the minimal evolving time of these systems. Also, Lloyd
have given out an attempting generalization of \ml in gravitational
systems. What's more, Zelinski and Zych have made modification on
\ml to evaluate some specific systems more precisely\cite{Zych}. In
this letter we will discuss these generalizations and modifications.
We will show that the generalized form of \ml proposed by Lloyd has
failed to apply on some systems, i.e. the radiation dominating era
of the universe. Lloyd's form is covariant but not the correct
covariant version for \mll. Thus we will give out our covariant
versions of \mll, which correctly evaluate all relativistic systems
and gravitational systems.
\\

This paper will be organized as follow.  in Sec. \ref{sec2}, we
revisit the \ml and its extension in frame of quantum mechanics. In
Sec. \ref{sec3}, the \ml on mixed state, composite system and
entanglement were discussed. It shows that \ml as an fundamental
limit on $\Delta t$ still works on these systems. In Sec.
\ref{sec4}, we focus on constructing a covariant form of \ml when
the special relativistic effect was taken into account. Since we
still don't have a complete quantum theory of gravity, we can't
obtain a similar theorem in general relativity from the fundamental
level. We analyze this problem in a semiclassical way in Sec.
\ref{sec5}, and get a similar conclusion with Sec. \ref{sec4}. After
that, the application of the formula of covariant quantum limits of
computation is discussed in Sec. \ref{sec6}. We discuss the close
relation among entropy bound, maximum information flow and ultimate
computational limit, it is found that the order of magnitude of
maximum logical operations $\#$ of a system during the amount of
time of communicating from one side to the other is no less than the
entropy of the system $S$, or the amount of information that the
system can store. Remarkably, this naturally gives a physical
interpretation of the inequality in the first set of hypotheses in
\cite{fmw}. Discussions and conclusions will be made in Sec.
\ref{sec7}.

\section{The \ml}\label{sec2}
  The \ml states a quantum system with energy expectation
value $E$ above its ground state energy, could transform from one
original state to its orthogonal state in the time $\Delta t$, where
$\Delta t \geqslant \pi \hbar / 2E$ \cite{ML}. Firstly, let us
consider an arbitrary quantum system could be decomposed as
superposition of its energy eigenstates,

\begin{equation} \label{eq:1}
|\Psi(0)\rangle=\sum_i C_i |\Psi^i_{E_i}\rangle ,
 \end{equation}

where $\Psi^i_{E_i}$ is the $i$th eigenstates of the system with
energy $E_i$. We assumed the system have a discrete spectrum and the
ground state energy is set zero. By quantum mechanics, after a time
interval $t$, the system will transform to,

\begin{equation} \label{eq:2}
|\Psi(t)\rangle=\sum_i C_i
e^{-\frac{i}{\hbar}E_it}|\Psi^i_{E_i}\rangle .
 \end{equation}

Then we set the inner product to be,

\begin{equation} \label{eq:3}
S(t)=\langle\Psi(0)|\Psi(t)\rangle=\sum_i |C_i|^2
e^{-\frac{i}{\hbar}E_it}.
 \end{equation}

In order to obtain the minimal value of evolving time, we use
inequality $\cos(x)\geqslant 1-2/\pi[x+\sin(x)]$, which is valid for
$x \geqslant 0$,Then,
\begin{equation} \label{eq:4}
\begin{aligned}
Re(S)&=\sum_i |C_i|^2 \cos{(\frac{i}{\hbar}E_it)}\\ &\geqslant
\sum_i |C_i|^2
[1-\frac{2}{\pi}(\frac{E_it}{\hbar}+\sin(\frac{E_it}{\hbar}))] \\
&=1-\frac{2E}{\pi\hbar}t-\frac{2}{\pi} Im(S).
\end{aligned}
 \end{equation}

The system reaches its orthogonal state while we set $S(\Delta
t)=0$.Then we finally obtain $\Delta t \geqslant \pi \hbar/2E$,
where $E$ is the expectation value of the system energy(The ground
state energy does not account for $t$, which only add a global phase
factor in equations). Some examples which saturate the inequality
have been found in \cite{ML}, so $\pi \hbar/2E$ gives a upper bound
of how fast can a system transform to its orthogonal state. This is
the main conclusion of \ml. If we use a different inequality such
as:
\begin{equation} \label{eq:5}
\cos(x)\geqslant
1-\frac{2}{\pi^\alpha}x^\alpha-\frac{2\alpha}{\pi}\sin(x),
 \end{equation}
where $\alpha>0$, then we obtain a more complicate form of the
estimation\cite{Zych}:
\begin{equation} \label{eq:6}
\Delta t\geqslant \frac{\pi\hbar}{2^{1/\alpha}
\langle(E-E_0)^{\alpha}\rangle^{1/\alpha} } , \ \  \alpha>0.
 \end{equation}

\begin{figure}

\centering \subfigure[ ]{\label{fig:subfig:a}
\includegraphics[scale=0.3]{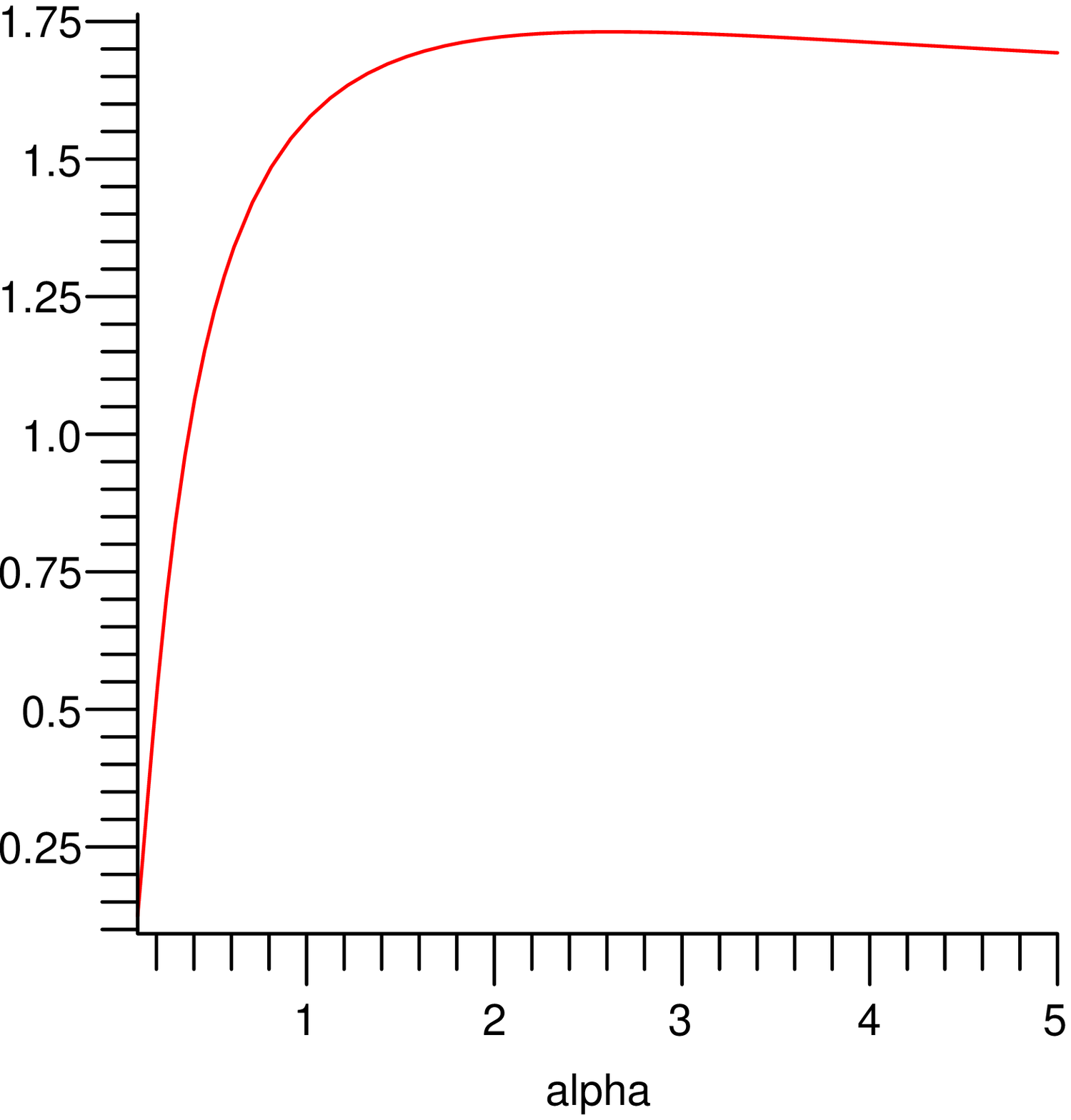}}
\hspace{0.1in} \subfigure[ ]{\label{fig:subfig:b}
\includegraphics[scale=0.3]{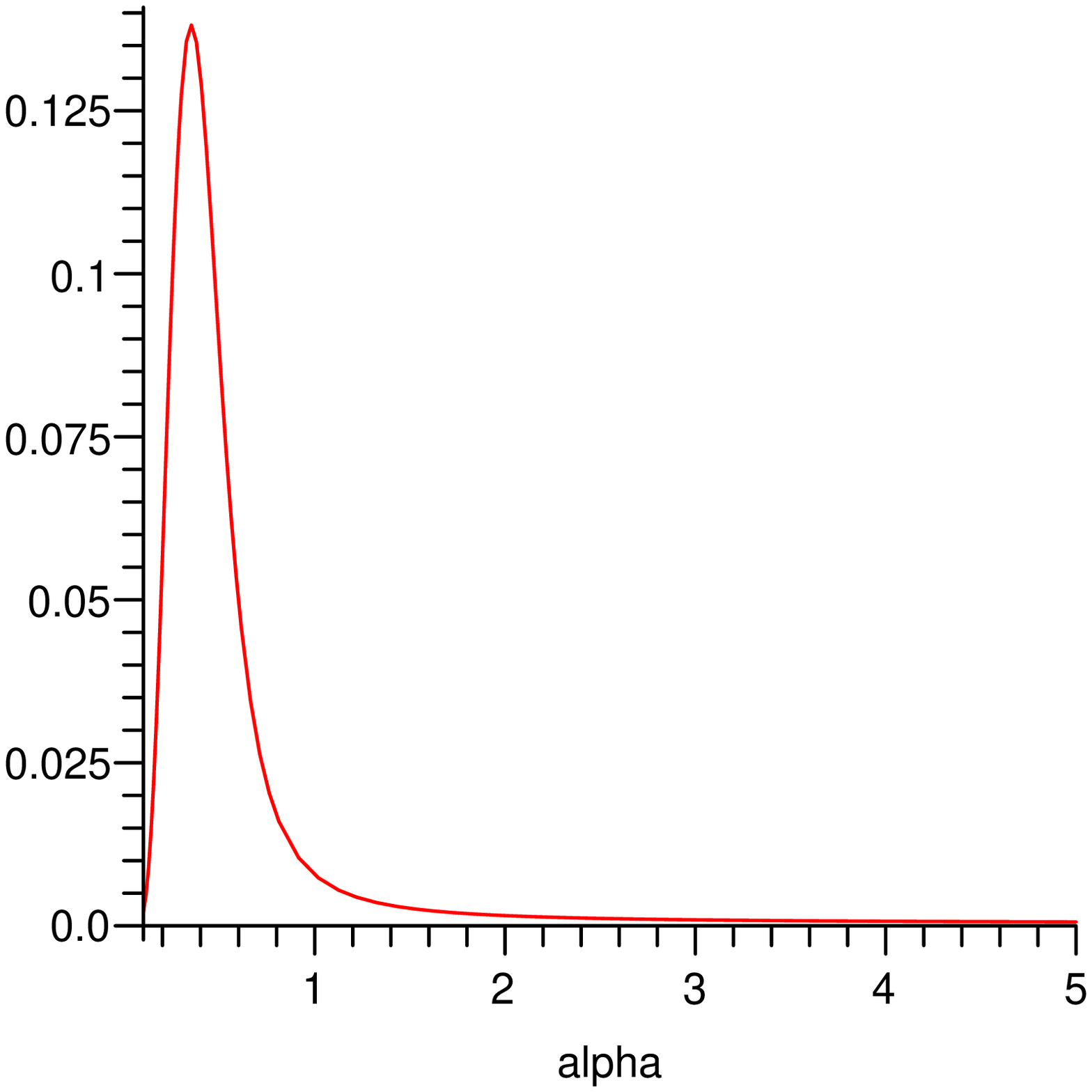}}
\hspace{0.1in} \subfigure[ ]{\label{fig:subfig:c}
\includegraphics[scale=0.3]{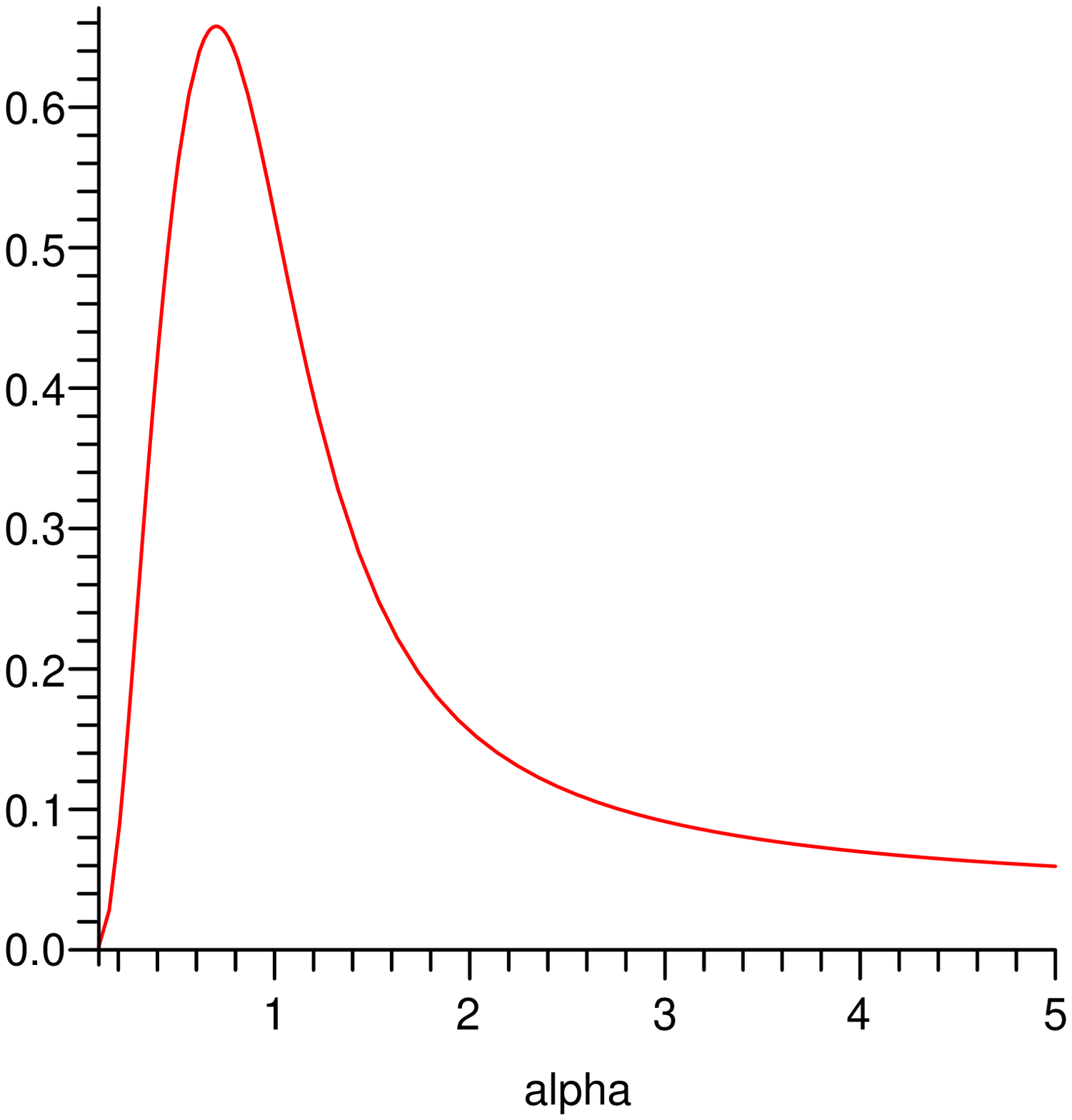}}
\caption{The value of $\pi\hbar/2^{1/\alpha}
\langle(E-E_0)^{\alpha}\rangle^{1/\alpha} $ for some specific
configuration of quantum states are plotted as the vertical axis. In
Figure (a) the initial state is
$|\Psi(0)\rangle=1/\sqrt{3}(|0\rangle+|1 \epsilon\rangle+|2
\epsilon\rangle)$. In Figure (b),
$|\Psi(0)\rangle=\sqrt{0.99}/\sqrt{2}(|0\rangle+|1
\epsilon\rangle)+0.1/\sqrt{2}(|10000 \epsilon\rangle+|10001
\epsilon\rangle)$. In Figure (c),
$|\Psi(0)\rangle=\sqrt{0.99}/\sqrt{2}(|0\rangle+|1
\epsilon\rangle)+0.1/\sqrt{2}(|100 \epsilon\rangle+|101
\epsilon\rangle)$. In all cases $\epsilon$ is set to be 1. It does
not show there is constant $\alpha$ as an universal parameter to
supply a tightest bound of minimal evolving time for all
configurations of quantum states.}
\end{figure}

Here $E_0$ is the ground state energy, and $\langle \rangle$ denotes
the expectation value on initial state. This new estimation $(\alpha
\ne 1)$ will get a higher bound than \ml$(\alpha = 1)$ on $\Delta t$
in some special case\cite{Zych}, but we are not able to obtain a new
bound with a constant $\alpha$, which limit $\Delta t$ more strictly
than \ml universally, as in Fig. 1. On the other hand, the physical
interpretation of the expression
$\langle(E-E_0)^{\alpha}\rangle^{1/\alpha}$ with $(\alpha \ne 1)$ is
obscure. The bound of Eq. (\ref{eq:6}) is more meaningful in
mathematics than in physics. However, \ml relates the minimal
evolving time to the average energy of the system, which offers a
relatively clear physical picture. As a result, we still perceive
\ml as a fine bound of $\Delta t$ in this letter, and all our
generalizations are based on \ml.

\section{\ml on mixed state, composite system and entanglement}\label{sec3}

In quantum mechanics, mixed states, composite systems and entangled
systems are always focused on as well as pure states, single and
non-entangled ones. Some previous work\cite{lloyd4,lloyd5} shows
that \ml as an fundamental limit on $\Delta t$ has its successful
applications on these systems. Consider a general state $\rho$, and
we assume $\rho$ can be decomposed with orthogonal energy
eigenstates,
\begin{equation} \label{eq:7}
\rho(0) = \sum_{i,j} C_{ij} |\Psi_i\rangle\langle\Psi_j|.
 \end{equation}
Here $\rho(0)$ can be viewed as a mixed state or a pure state,
depending what value $C_{ij}$ takes. Time translation denoted by an
unitary operator $\hat{U}(t)=e^{-i\hat{H}t/\hbar}$ transform
$\rho(0)$ to $\rho(t)$,
\begin{equation} \label{eq:8}
\rho(t) =\hat{U}(t)\rho(0)\hat{U}^\dagger (t)= \sum_{i,j} C_{ij}
e^{-i\hat{H}t/\hbar}|\Psi_i\rangle\langle\Psi_j|e^{i\hat{H}t/\hbar}.
 \end{equation}
A widely used function fidelity $F(\rho(0),\rho(t))$ is devoted to
discriminate orthogonality on mixed states. When we set $F=0$ to
confirm the orthogonality, it is easy to obtain:
\begin{equation} \label{eq:9}
F(\rho(0),\rho(t))=Tr
\sqrt{\rho^{1/2}(0)\rho(t)\rho^{1/2}(0)}=\sum_i C_{ii}
e^{-iE_{i}t/\hbar}
 \end{equation}
$E_i$ is energy value of $|\Psi_i\rangle$(Here the Hamiltonian is
assumed to be diagonal). We can easily see that fidelity coincide
with the inner product between initial and final states, while
$\rho$ is specialized to a pure state. Further deduction gives out
$\Delta t \geqslant \pi \hbar/2E$, and $E=Tr (\hat{H} \rho)$ is the
energy expectation value of the mixed state. Noting that we have not
taken interacting Hamiltonian in our consideration yet. Directly,
this generalization on mixed state lead us to investigate the self
consistency of \ml on all kinds of composite quantum systems. For
example, a mixed state $\rho$ can be composed by several pure
states($\rho=\sum_i p_i \rho_i$), or, a pure state can be
represented as directed product of several pure states, etc. A
question is asked, what is relationship between the minimal time to
orthogonal state $\Delta t$ of the system and $\Delta t_i$ of the
$i$th subsystem? It is answered that in the latter case, $\Delta t$
is no longer limited by $\pi \hbar/2E$, but,
\begin{equation} \label{eq:10}
\Delta t = max (\frac{\pi\hbar}{2E_i}).
 \end{equation}
Here $E_i$ is the energy expectation value of the $i$th subsystem,
unless one of the subsystems has all the energy of the
system\cite{lloyd4}. It does mean a composite system may reach its
orthogonal state with longer time than single ones with the same
average energy. However, the total operations performed by the
system should independent of its architecture\cite{lloyd4}. In one
second, a system with average energy $E$ would perform as most
$1/\Delta t=2E/\pi \hbar$ operations, which is the exact summation
of maximum operations performed by the subsystems($\sum_i 1/\Delta
t_i=\sum_i 2E_i/\pi \hbar=2E/\pi \hbar$). This is another advantage
of \mll, the operation rate of any system/subsystem limited by \ml
is proportional to its energy expectation value, which is a fine
extensive variable. In contrast, the operation rate limited by
energy spread $\Delta E$ \cite{lloyd4,lloyd1} or Eq. (\ref{eq:6}) is
not additive generally, which could cause some severe problem.
\\

So far we have discussed the minimal time to orthogonal state of a
non-entangled system with a free Hamiltonian. Although the entangled
case or interacting case of \ml is difficult to formulated, some
testing calculation \cite{lloyd4} implies a composite system with
entanglement-generating Hamiltonian are capable to operate with a
rate closer to the \ml bound than without it. The differences are
dependent on the dominance of interacting, while \ml is still valid
in these cases.

\section{A special relativistic form of \ml}\label{sec4}

\ml enable us to evaluate how many operations or events can be
discriminated at most of a 4-D space-time system, within a section
of its history. The result should be invariant under any physical
transformation of the observer. But apparently, the above form of
the bound is not invariant in relativity. Assuming a system
$\mathcal{S}$ with average energy $E$ could be reported with $\#
\leqslant 2E\tau/{\pi \hbar}$ operations at most in time $\tau$, in
a rest frame $O$. Now there is another observer in frame $O^\prime$,
watching $\mathcal{S}$ is moving along the $x$ axis with velocity
$v$, with an average energy $E\gamma$ for a time interval $\tau
\gamma$ in the same history, where $\gamma =1/ \sqrt{1-v^2}$. The
new limitation of operations in frame $O^\prime$ is $\#' \leqslant
2E\gamma^2\tau/{\pi \hbar}$. This paradox does not even exist in
special relativity. Accordingly, a moving clock ticks slower than a
same static clock with a rate $\gamma$ for time dilatation effect.
And the quantities of the ticks or operations detected from the
clock is independent of observers for causality. Margolus and
Levitin has already give a testing form\cite{ML} of the bound of
operation quantities detected in $O^\prime$, in a special relativity
view, which is $\# \leqslant 2px/{\pi \hbar}$. The inner product of
4-momentum $p^\mu$ and 4-translation $x^\mu$ guarantees Lorentz
invariance of the bound. And this result is easily achieved by
replacing quantum field theory of quantum mechanics in deduction.
For simplicity we focus on a state $\Psi$ composed by massive
particles\cite{weinberg},
\begin{equation} \label{eq:11}
\Psi=\Sigma_i C_i \Psi^i_{p_1,\sigma_1,n_1;p_2,\sigma_2,n_2;...}.
 \end{equation}
The indexes mean the $i$-th eigenstate of $\Psi$ is occupied by an
$n_1$-th kind of particle with 4-momentum $p_1$ and spin $\sigma_1$
and so on. In quantum field theory, a time translation $(t,0,0,0)$
is perceived as a trivial Lorentz transformation of the coordinates
by $(-t,0,0,0)$, as
\begin{equation} \label{eq:12}
U(1,-t)\Psi=\Sigma_i e^{-i(p_1^0+p_2^0+...)t} C_i
\Psi^i_{p1,\sigma1,n1;p2,\sigma2,n2;...}.
 \end{equation}
With a Lorentz boost, the state $\Psi$ transforms to,
\begin{equation} \label{eq:13}
U(\Lambda,0)\Psi=\Sigma_i C_i \times \sqrt{\frac{(\Lambda
p_1)^0...}{p_1^0...}} \Sigma_{\sigma_1\prime,\sigma_2\prime,...}
D^{(j_1)}_{\sigma_1^{\prime}\sigma_1}(W(\Lambda,p_1))...
\Psi^i_{\Lambda p_1,\sigma^\prime_1,n_1;\Lambda p_2,\sigma^\prime
_2,n_2;...}.
 \end{equation}

Then the time translation $(t,0,0,0)$ becomes spacetime translation
after the boost,

\begin{equation} \label{eq:14}
\begin{aligned}
& U(\Lambda,0)U(1,-t)\Psi= U(\Lambda,-\Lambda t)\Psi
\\
&=\Sigma_i C_i \times e^{i(\Lambda t)_\mu \Lambda(p_1+p_2+...)^\mu}
\times \sqrt{\frac{(\Lambda p_1)^0...}{p_1^0...}}
\Sigma_{\sigma_1\prime,\sigma_2\prime,...}
D^{(j_1)}_{\sigma_1^{\prime}\sigma_1}(W(\Lambda,p_1))...
\Psi^i_{\Lambda p_1,\sigma^\prime_1,n_1;\Lambda p_2,\sigma^\prime
_2,n_2;...}.
\end{aligned}
 \end{equation}
Here, $\Lambda$ is a Lorentz transformation to the coordinate, $W$
is the Lorentz transformation belonging to the little group, and $D$
furnish a representation of the little group. More details can be
found in \cite{weinberg}. Then the form of inner product between the
two states becomes relevant to the Lorentz boost,
\begin{equation} \label{eq:15}
\begin{aligned}
S=\Psi^* U(1,-t)\Psi &=\Sigma_{i} |C_i|^2 e^{-i(p_1^0+p_2^0+...)t}
\\&=(U(\Lambda,0)\Psi)^*U(\Lambda,-\Lambda t)\Psi
\\&=\Sigma_{i} |C_i|^2
e^{-i(p^{\prime}_1+p^{\prime}_2+...)x^{\prime}}.
\end{aligned}
 \end{equation}
We still use inequality $\cos(x)\geqslant 1-2/\pi[x+\sin(x)]$ to
obtain the special relativistic form of \mll,
\begin{equation} \label{eq:16}
\begin{aligned}
Re(S)&=\sum_i |C_i|^2 \cos{(\frac{i}{\hbar}(p^\prime_1+...)x^\prime)}\\
&\geqslant \sum_i |C_i|^2
[1-\frac{2}{\pi}(\frac{(p^\prime_1+...)x^\prime)}{\hbar}+\sin(\frac{(p^\prime_1+...)x^\prime)}{\hbar}))] \\
&=1-\frac{2p^\prime x^\prime}{\pi\hbar}-\frac{2}{\pi} Im(S).
\end{aligned}
 \end{equation}
Where $p^\prime_i=\Lambda p_i$, $x^\prime=\Lambda
t=(1,\vec{v})t\gamma$, $\gamma=1/\sqrt{1-\vec{v}^2}$ and
$p^\prime=\sum_i |C_i|^2 (p^\prime_1+p^\prime_2+...)$ is the
expectation value of 4-momentum of the system. A quick analysis
gives the bound of minimal evolving time of orthogonal state of the
system,
\begin{equation} \label{eq:17}
\Delta t^\prime \geqslant \frac{\pi \hbar}{2p^\prime \cdot
(1,\vec{v})}.
\end{equation}
The physical picture of this inequality is confusing. Because time
is not a Lorentz invariant variable, it is not wise for us to seek a
relativistic invariant form of minimal evolving time between
orthogonal states of the system. Although a system evolving for a
time interval $\tau$ in the rest frame $O$, could be described
undergoing a dilated time $\tau \gamma$ in the moving frame
$O^\prime$, the bound of quantities of operations or events in the
same history is still invariant,
\begin{equation} \label{eq:18}
\#_{bound}=\frac{2E\tau}{\pi\hbar}=\#_{bound}^\prime=\frac{\tau
\gamma}{\Delta t^\prime}=\frac{2p x}{\pi \hbar}.
\end{equation}
Where we denote $p^\prime$ and $\tau \gamma (1,\vec{v})$ by $p$ and
$x$ for simplification, and we denote the upper bound of quantities
of operations by $\#_{bound}$. This result is inevitable, since $px$
is an direct covariant representation of $E\tau$ in the rest frame
$O$. If the 3-components of $p$ are all zero----it means the system
is relative static to observer----the upper bound $\#_{bound}$
degenerates to the original form $E\tau$. It shows \ml still works
on systems static to observer, and evaluating a covariant bound of
$\#$ is more practical than a covariant bound of $\Delta t$. Hence
 we should seek a generalized form of \ml to give the bound of $\#$
 of a system other than $\Delta t$.
\\

As long as the system $\mathcal{S}$ is simple, it is possible to get
the 4-momentum $p$ and 4-translation $x$ of the system, and to make
an inner product. However in practice, it is hard to define
4-momentum and 4-translation of some complex system such as fluids
or radiations. In most theoretical analysis the enery-momentum
tensor $T_{ab}$ is more general than 4-momentum, so a generalized
form of \ml should take $T_{ab}$ as an component of estimation of
the bound. Apparently, an observer with unit timelike 4-velocity
$u^a$ in Minkowski spacetime will obtain an energy current density
$J_a = -T_{ab}u^b$\cite{wald}, as measuring a fluid with energy
momentum tensor $T_{ab}$, and the current density satisfies
$\partial^aJ_a = 0$. By using Gauss's law, we can get the
conservation of energy. Thus we can straightforwardly replace $p_a$
in Eq. (\ref{eq:18}) by $\int dV T_{ab} u^b$(here a constant $c^2$
is omitted) and then we get,
\begin{equation} \label{eq:19}
 \# \leqslant
\frac{2}{\pi\hbar} \int T_{ab} u^a \frac{dx^b}{d\tau} dV d\tau=
\frac{2}{\pi\hbar} \int T_{ab} u^a u^b dV d\tau .
 \end{equation}

The equivalence between Eq. (\ref{eq:19}) and \ml can be proved by
setting $T_{ab}=p_a p_b / E \delta^3(\vec{x}-\vec{x}(\tau))$
\cite{weinberg2}, noting that a 3-volume contracts by a factor
$\gamma$ under boost of Lorentz transformation. Since $x$ is
4-translation of the system, $u^a=dx^a/d\tau$ is not an arbitrary
observer's 4-velocity, but the 4-velocities of observers at rest
with respect to the system. This ensures the bound of the system
with same history is not depending on different observers. The
covariance of the generalized form is obvious. $T_{ab} u^a u^b$ is a
scalar and $dV d\tau$ is an invariant 4-volume unit in a flat
background. The operational capacity of a system is now determined
by $T_{ab} u^a u^b$ and the most quantities of operations or events
can be detected is proportional to the size of the 4-volume of the
system. It must be emphasized that the non-negative $\#$ requires
the system obeying the weak energy condition,
\begin{equation} \label{eq:20}
T_{ab} u^a u^b \geqslant 0.
 \end{equation}
Though in some theory, system with matter violating weak energy
condition is not unstable and is able to evolve from its initial
state to an orthogonal state to perform operations, the validity of
\ml on these systems is still to be evaluated carefully. We assert
that our estimations of upper bound of quantities of operations are
following the weak energy condition.
\\

The generalized form of \ml has a direct application on cosmology.
Consider a flat universe with constant $\omega$ above -1, which
describes the massive dust dominant era and radiation dominant era
of the universe. We have Friedmann equations with a flat metric:
\begin{equation} \label{eq:21}
 H^2=\frac{8 \pi G}{3} \rho ,
 \end{equation}
\begin{equation} \label{eq:22}
 \dot{\rho}+3H\rho(1+\omega)=0 ,
 \end{equation}
\begin{equation} \label{eq:23}
 ds^2=-dt^2+a^2(t)[dr^2+r^2d\Omega].
 \end{equation}
Here $T_{ab}=(\rho+p)u_a u_b+p g_{ab}$ is the energy-momentum tensor
of the dominating matter, $p$ is pressure density, $\rho$ is energy
density and we have $p=\omega \rho$. Resolving Friedmann equations,
we obtain,
\begin{equation} \label{eq:24}
 H=\frac{2}{3(1+\omega)(t-t_0)}.
 \end{equation}
We will omit the constant $t_0$ for following discussion.
Substituting Eq. (\ref{eq:21}), Eq. (\ref{eq:24}) and the
representation of $T_{ab}$ into Eq. (\ref{eq:19}), and take the
apparent horizon $l_H=c/H$ as the physical bound of our universe, we
obtain,
\begin{equation} \label{eq:25}
 \# \leqslant \frac{9c^5t^2(1+\omega)}{4\pi \hbar G}.
 \end{equation}
The result has the same order of Lloyd's result $\# \approx
c^5t^2/G\hbar$\cite{lloyd2}. According to our result, the bounds of
the quantities of events detected in our universe vary from
different expanding history. This is different with Lloyd's result,
since his deduction has omitted the expanding details of the
universe. The result depending on $\omega$ can be explained that
different expanding history dilute the energy density of the
universe with different rate, while the expanding horizon contains
more operational units from outside with different rate. Then the
form of upper bound of the total events in the observed universe is
different among different expanding eras, while it always scales
similar to the area of the horizon. It is consistent with and
complementary to the Bekenstein bound\cite{lloyd3}. Any appliance of
Eq. (\ref{eq:19}) on the $\omega=-1$ matter dominating era is not
successful, e.g. the cosmological constant. Since \ml is discussing
average energy above the ground state, it is not reasonable applying
\ml on cosmological constant, which could be interpreted to vacuum
energy.

\section{A general relativistic form of \ml}\label{sec5}

In the special relativity, we find that the upper bound of
operations is associate with a conserved quantity of energy.
However, in general relativity, there is not a notion of the local
stress-energy tensor of the gravitational field. Nevertheless, for
asymptotically flat spacetimes, conserved quantity of energy
associated with asymptotic symmetries has been well defined at
spatial and null infinity\cite{wald2}. Since there are lots of
approaches to define conserved energy, we assert that our result may
not be unique in the most general contexts.
\\

In general relativity, the energy properties of matter also
represented by a energy momentum tensor $T_{ab}$, and the energy
momentum current is still:
\begin{equation}
J_a =-T_{ab}u^b.
\end{equation}
The local energy density $\rho$ of matter as measured by a given
observer with unit timelike four-velocity $u^a$ is well defined
by\cite{wald}:
\begin{equation}\label{rho}
\rho =T_{ab}u^au^b.
\end{equation}
So, locally the energy in a cell with spacial volume $\Delta V$ is
$T_{ab}u^au^b\Delta V$, the maximum logical operations $\Delta
\#_{bound}$ in the cell within proper time $\Delta\tau$ of the
observer is:
\begin{equation}\label{ops}
\Delta \#_{bound} = \frac{2}{\pi\hbar}T_{ab}u^au^b\Delta\tau\Delta
V.
\end{equation}
The form is very similar with Eq. (\ref{eq:19}). One would expect
Eq. (\ref{eq:19}) may be still valid in general relativisty.
However, the condition $\nabla^aT_{ab}=0$ does not lead to a global
conservation law in general relativity($\nabla^aJ_a$ is generally
non-zero, not like in special relativity). There is no conserved
total energy in general relativity, unless there is a Killing vector
$\xi^a$ in the spacetime, i.e.,$\nabla_{[a}\xi_{b]}=0$. In this case
$\nabla^a J_a =\nabla^a (T_{ab}\xi^b) =0$, and the conserved total
energy associated with the Killing field $\xi^a$ on a spacelike
Cauchy hypersurface $\Sigma$ can be defined as,
\begin{equation}\label{energy}
    E = \int_\Sigma T_{ab}n^a\xi^b d\Sigma,
\end{equation}
where $n^a$ denotes the unit normal vector to $\Sigma$. Here the
spacetime $(M,g_{ab})$ have been foliated by Cauchy hypersurface
$\Sigma_t$ which is parameterized by a global time function $t$,
then the metric can be write in ADM form,
\begin{equation}\label{metic}
    ds^2 = g_{ab}dx^adx^b = -N^2dt^2+h_{ij}(dx^i+N^idt)(dx^j+N^jdt).
\end{equation}
$N$ is called the {\it lapse function} which measures the difference
between the coordinate time $t$ and proper time $\tau$ on curves
normal to the Cauchy hypersurfaces $\Sigma_t$, $N^i$ is called the
{\it shift vector} which measures the difference between a spatial
point $p$ and the point one would reach if instead of following $p$
from one Cauchy hypersurface to the next one followed a curve
tangent to the normal $n^a$, and $h_{ij}$ is the intrinsic metric
induced on the hypersurface by $g_{ab}$, The timelike normal $n^a$
defines the local flow of time in $M$.
\\

The energy defined in Eq. (\ref{energy}) is just the total energy on
$\Sigma$ observed by an observer with 4-velocity $\xi^a$
, the maximum logical operations of $\Sigma$ over proper time $\int
dt$ is bounded by
\begin{equation}\label{ops2}
    \# \leqslant \frac{2}{\pi\hbar}\int T_{ab}n^a\xi^b d\Sigma dt.
\end{equation}
Alternatively, from the definition of energy density $\rho$ in Eq.
(\ref{rho}), we suspect the maximal logical operations the fluid
over time $\int dt$ would have the form,
\begin{equation}\label{energy2}
    \# \leqslant \frac{2}{\pi\hbar}\int N\,T_{ab}u^au^b d\Sigma dt.
\end{equation}
This definition of energy coincides with the definition in
(\ref{energy}) when the Killing vector $\xi^a$ satisfies,
\begin{equation}\label{relation}
    \xi^a = N u^a \ and\  u^a=n^a.
\end{equation}
This means Eq. (\ref{energy2}) is valid when the spacetime manifold
have symmetries which can generate time killing vectors normal to
$\Sigma$. We can write $d\mathcal{V}=Nd\Sigma dt$ as the proper
volume element of spacetime 4-volume $\mathcal{V}$, then
\begin{equation}\label{ops3}
    \# \leqslant \frac{2}{\pi\hbar}\int N\,T_{ab}u^au^b d\Sigma dt=\frac{2}{\pi\hbar}\int_V T_{ab}u^au^b
    d\mathcal{V},
\end{equation}
Obviously, this form of \ml is covariant and consistent with Eq.
(\ref{eq:19}) in special relativity. So far we know little about
quantum gravity, we are not able to give a deduction of the
generalized form of \ml from the fundamental level. However, the
picture of the generalized form is proved to be compatible with
quantum mechanics and general relativity.
\\

Consider a fluid with energy momentum tensor $T_{ab}=(\rho+p)u_a
u_b+p g_{ab}$ spreads in a spacetime region with gravitational field
$g_{ab}$. We assume the distribution of the fluid is static and
spherical symmetric, and the size of the system is $l$, the bound of
the integral is omitted for simplification. Then the upper bound of
quantities of operations is represented by,
\begin{equation}\label{eq:35}
\# \leqslant \frac{2}{\pi\hbar} \int \rho(r) r^2 dr d\Omega dt.
\end{equation}
Here $\Omega$ is the area of a unit 2-sphere, and $d\mathcal{V}=dr
d\Omega dt$. It seems there is no difference between Eq.
(\ref{eq:35}) and the special relativistic form. However, the
gravitational effect has already been included in Eq. (\ref{eq:35}).
The \ml relates the minimal evolving time to the average energy of
the system, while the average energy here should be the sum of
energy not gained from gravity and energy gained from gravity. Here
the energy not gained from gravity is $E_{ng} = \int
1/\sqrt{-g_{00}} \rho(r) dr^2 d\Omega$,and the energy gained from
gravity is $E_g = - \int \rho(r)[1/\sqrt{-g_{00}}-1]r^2 dr
d\Omega$\cite{gravitation}. $E_g$ is below zero, and is lowering
down the total average energy. This implicates that the system
performs operations slower in gravitational field. While since a
system in gravitation field has the same average energy with an
other system without considering gravity, as measured by an observer
from infinity, they have the same bound of quantities of operations,
even one of them performs slower. This can be explained from the
other aspect as following. An operation unit of the system takes
$\Delta T$ at infinity to reach its orthogonal state. Because of the
time dilation effect of general relativity, if we put the unit into
gravitational field, the minimal evolving time will expands to
$\Delta t=1/\sqrt{-g_{00}}\Delta T$ as measured by observers at
infinity. However, the physical size of system in gravitational
field is $\sqrt{g_{11}}$ times larger than the size as measured by
observers at infinity. If the outside of the static spherical system
is vacuum, we can obtain $-g_{00}(r)g_{11}(r)=1$. As a result, the
system in gravitational field could maximally perform as many
operations as the system with the same average energy out of
gravitational field. Negative binding energy keeps the bound fixed
although the operation units are slowed and the amount of operation
units become large. This is the similar mechanism with the analysis
entropy bound of gravitational system in monster-like
state\cite{monster}. Also, the generalized form of \ml gives a
complimentary explanation to the accuracy of spacial measurement
$\delta l \sim l^{\frac{1}{3}}$\cite{lloyd2,Ng,Xiao}. Since the
spacetime is distorted by gravity, the operations and events are not
spreading homogeneously between spacial section and temporal section
of the 4-volume. In the characteristic time of communication $l/c$,
each operation unit of the system can perform only of order one
operation. Then the relax of temporal resolution from $\delta t \sim
l^{\frac{1}{2}}$ (non-gravitational case) to $\delta t \sim l$
corresponds to the enhancement of spacial resolution from $\delta l
\sim l^{\frac{1}{2}}$(non-gravitational case) to $\delta l \sim
l^{\frac{1}{3}}$. While in both cases the maximal quantities of
events contained in a 4-volume $l^4$ are both scaling as $l^2$.
\\

Lloyd has given another covariant version of \ml \cite{lloyd3},
\begin{equation}\label{eq:36}
\# \leqslant -\frac{2}{\pi\hbar} \int T_{ab}g^{ab}d\mathcal{V},
\end{equation}
where the inequality is set in a rest frame with an inertial clock.
For the fluid we have discussed above, the covariant quantity
$-T_{ab}g^{ab}$ can be simplified to $\rho-3p$. Then Eq.
(\ref{eq:36}) is equivalent to Eq. (\ref{eq:35}) when the fluid is
composed by massive dust($p(r)=0$). For fluids ($p(r) \ne 0$), Eq.
(\ref{eq:36}) gets different results with Eq. (\ref{eq:35}). If we
let the fluid to be radiations which is the dominating component of
the universe in some era. The characteristic feature of radiations
is $p=1/3 \rho$, which makes the maximal operations of the
radiations obtained by Eq. (\ref{eq:36}) to be 0. It means the
quantum units in radiations will never evolve to their orthogonal
states. This result can be avoided if we take Eq. (\ref{eq:35}) as
the covariant version of \mll, since the pressure density of the
fluid is not contained apparently in the expression.
\section{entropy bound, maximal information flow and computational limit}\label{sec6}

The entropy bound measures the capability of how much the
information can be coded in a system. The maximum entropy flow rate
associated with maximum speed of information transfer, is said the
limit of communication. And computational limit reflects the ability
that the system can deal with information or produce new
information. We will find those three bounds are compatible with
each other. First, we will discuss Bekenstein system. With our
covariant form of \mll, some result can be generalized to Bousso
system. Bekenstein system and Bousso system will be defined below.
For convenience, Planck units will be used in this section, i.e., $c
= G = \hbar = k = 1$, where $c$ is the speed of light, $G$ is Newton
constant, $\hbar$ is Planck constant, and $k$ is Boltzmann constant.

\subsection{In Bekenstein System}

For any complete, weakly self-gravitating, isolated matter system in
ordinary asymptotically flat space,
Bekenstein\cite{Bekenstein,bekenstein2} has argued that the
Generalized Second Law of thermodynamics(GSL) implies a bound to the
entropy-to-energy ratio
\begin{equation}\label{ratio}
    \frac{S}{E}\leqslant 2\pi R,
\end{equation}
here,$E$ is the total energy of the matter system, $R$ is the radius
of the smallest sphere that fits around the matter system. The bound
has been explicitly shown to hold in wide classes of equilibrium
systems\cite{sb}, we will call this kind of system that satisfies
the conditions for the application of Benkenstein's bound as {\it
Bekenstein system}\cite{bousso1}.
\\

Since energy is well defined in {\it Benkenstein system}, it can be
viewed as a computational system.Then, according to the \mll, the
maximal logical operations that the computer can perform during the
time $t_{com}=2R$ is
\begin{equation}\label{opcom}
\#_{bound}=2Et_{com}/\pi =4ER/\pi,
\end{equation}
$t_{com}$ is just the time for sending a message from one side of
the system to the other. Using the Bekenstein bound, i.e., Eq.
(\ref{ratio}), we have,
\begin{equation}\label{ratio2}
    \frac{\#_{bound}}{S} =\frac{4ER}{\pi S}\geqslant \frac{2}{\pi^2},
\end{equation}
which is a lower bound of $\#_{bound}$-to-$S$ ratio. We find
$\#_{bound}$ and $S$ have the same order, which means the order of
magnitude of maximum logical operations of a {\it Bekenstein system}
$\#_{bound}$ is no less than the entropy of the system $S$ during an
amount of time for communicating from one side to the other. In
fact, Lloyd\cite{lloyd1} states that the measurement of the degree
of parallelization in an ultimate computer is $t_{com}/t_{flip}$,
where $t_{flip}$ is the characteristic time for flipping a bit in
the system. For a {\it Bekenstein system} $t_{com}/t_{flip}
\geqslant 2/\pi^2$, which means the operation of a {\it Bekenstein
system} can not be totally serial. Here, we just reinterpret in
another way.
\\

Why the operation of a computational system is always somewhat
parallel? The answer is we can't transfer information at an
arbitrary speed. It has been broadly discussed\cite{bekenstein5}
that the information flow rate is bounded by the configuration of
the system. Bekenstein presents a bound on information flow
$\dot{I}$\cite{bekenstein3},
\begin{equation}\label{sflow}
  \dot{I}\leqslant \frac{\pi E}{\ln 2}.
\end{equation}
It is a reinterpretation of the proposition in \cite{bekenstein4} as
he said. This bound is consistent with previous work by Bremermann
and Pendry etc\cite{bekenstein5}. How much information can be
transferred during the time $t_{com}=2R$ ? By using (\ref{sflow}),
it is easy to see that the information $I\leqslant 2\pi ER/\ln 2$,
and the entropy taken by the information is $S\leqslant 2\pi ER$.
This means all the information stored in a system can be transfered
in a time it takes communicate from one side to the other if we can
make the communication channel reach it's capacity limitation. It is
also worth to note that equation (\ref{ratio2}) tell us the ultimate
computer can produce new information as much as the ultimate
communication channel can transfer, so that any computer will always
be somewhat parallel.

\subsection{In Bousso System}

The form of Bekenstein's entropy bound is unapparently covariant.
Bousso has already given out an apparently covariant entropy bound
for gravitational systems. The system satisfying Bousso's entropy
bound is called {\it Bousso system}. We find that the correspond
relation of Eq. (\ref{ratio2}) in {\it Bousso system} is still
valid. We will discuss the validity of the $\#_{bound}$-to-$S$ ratio
lower bound in gravitational collapse system and cosmological
models.

\subsubsection{Testing the bound in gravitational collapse}

In a gravitational collapsing system, the internal mass distribution
is not homogeneous. Some regions are over-dense, and the light-rays
which we used to send signals from one side to the other side are
deflected. The signals might take intricate, long-winded path across
the gravitational collapsing system\cite{bousso1}. This means it
takes more time to send the messages across the system, and more
logical operations should be taken into account in Eq.
(\ref{ratio2}). It is quite plausible that this mechanism will
protect the $\#_{bound}$-to-$S$ ratio lower bound from being
violated.

\subsubsection{Testing the bound in lightsheet}

In \cite{bousso1}, Bousso shows the covariant entropy bound resolved
the limitation in using Fischler-Susskind bound\cite{fs} to closed
FRW universe, and take it as a strong evidence for the validity of
his conjecture. All the quantitative calculation were taken under
the assumption that the entropy in the universe can be described by
local entropy density $s$ and the matter content can be described by
a perfect fluid, with stress tensor $T^a\!_b=diag(-\rho,p,p,p)$. It
is quite plausible that for this kind of matter the hypothesis in
\cite{fmw},
\begin{equation}\label{hypothesis}
    \pi (\lambda_\infty -\lambda) T_{ab}k^ak^b \geqslant
    \left|s_ak^a\right|,
\end{equation}
will be satisfied. This inequality is sufficient to prove the
generalized Bousso Entropy bound and all the result in sec.
\ref{sec4} of \cite{bousso1} is naturally.
\\

If we also make the assumption that there is an entropy flux
4-vector $s^a$ associated with each lightsheet $L$ whose integral
over $L$ is the entropy flux through $L$ as in \cite{fmw}. Using the
same notation in \cite{fmw}, the entropy flux through $L$ is given
by
\begin{equation}\label{sflux}
    S_L=\int^{B^\prime}_B d^2x\sqrt{det
    h_{\Gamma\Lambda}(x)}\int^{\lambda_\infty(x)}_0 d\lambda
    \,s(\lambda)\mathcal{A}(\lambda).
\end{equation}
Here $x\equiv(x^1, x^2)=x^\Gamma$ is a coordinate system on the
2-surface $B$, $h_{\Gamma\Lambda}(x)$ is the induced 2-metric on
$B$, $\lambda$ is an affine parameter, $k^a=(d/d\lambda)^a$ is
tangent to each geodesic, and $\lambda_\infty(x)$ is the value of
affine parameter at the endpoint of the generator which starts at
$x$, $s\equiv -s_ak^a$ is non-negative for future directed timelike
or null $s^a$, $\mathcal{A}(\lambda)\equiv exp[\int_0^\lambda
d\bar{\lambda}\theta(\bar{\lambda})]$ is an area-decreasing factor
in the given geodesic with $\theta$ interpreted as the expansion.
\\

In the same way, the energy flux through $L$ is
\begin{equation}\label{eflux}
    E_L=\int^{B^\prime}_B d^2x\sqrt{det
    h_{\Gamma\Lambda}(x)}\int^{\lambda_\infty(x)}_0 d\lambda
    \,\rho (\lambda)\mathcal{A}(\lambda),
\end{equation}
where $\rho=-J_ak^a=T_{ab}k^ak^b\geq 0$ if we assume the null
convergence condition.
\\

Noting that the amount of time for sending a signal from $B$ to
$B^\prime$ is just $\lambda_\infty -\lambda$ in Planck unit. Then,
by using equation (\ref{hypothesis})(\ref{sflux}) and (\ref{eflux}),
the $\#_{bound}$-to-$S$ ratio lower bound hold on any lightsheet
followed immediately, i. e.
\begin{equation}\label{ratio3}
\frac{\#_{bound}}{S} =\frac{2E_L(\lambda_\infty -\lambda)}{\pi
S_L}\geqslant \frac{2}{\pi^2}.
\end{equation}
Now the physical meaning of the hypothesis is specific, and the
computational limits and the entropy bound is related to each other
very closely.
\section{Conclusions and discussion}\label{sec7}

In this paper, we revised \ml and compared it with other form of
limitation of minimal evolving time. We stated that \ml has a clear
physical picture and has successful applications on composite
systems, mixed state and entanglement systems. Following the spirit
of \ml, we have presented covariant versions of \ml respectively in
special relativity and general relativity. The covariant version in
special relativity is deduced analytically, while the version in
general relativity is still tentative but reasonable. After that we
give some applications of the versions of \ml on cosmology and
quantum gravitational systems. In the end, we discussed the relation
between entropy bound, maximal information flow and computational
limit. All the applications were consistent with previous works, and
presented more details and physical pictures on those topics.
\\

Although the covariant version of \ml in general relativity has its
successful applications, we still need estimations to evaluate its
validity more completely. So far the deduction shows the objects of
relativistic version of \ml are configured on space-like surfaces,
excluding null-surfaces. Because anything observed by an
relativistic observer should be configured on the null-surfaces
called light-cones, the deduction of \ml needs to be extended to
null-surfaces.

\section{Acknowledgements} We would like to thank Q. Ma for useful
discussions. The work is supported in part by the NNSF of China
Grant No. 90503009, No. 10775116, and 973 Program Grant No.
2005CB724508.


\end{document}